\documentclass[12pt]{article}
\usepackage{mathtext}
\usepackage{graphicx}
\usepackage[T2A]{fontenc}
\usepackage[utf8]{inputenc}
\usepackage{amsfonts}
\usepackage{amssymb}
\usepackage{amsmath}
\usepackage{pgfplots}
\usepackage{braket}
\usepackage{tikz}
\usepackage{float}
\usepackage{multicol}
\newcommand{\w}{\omega}
\newcommand{\s}{\sigma}

\begin{document}
    \title{Collapse of dark states in Tavis-Cummings model}
    
    \author{Vitaliy Afanasyev$^{1}$, Chen Ran$^{1}$, Yuri Ozhigov$^{1,2}$, You Jiangchuan$^1$ \\
    {\it 
    1.Lomonosov Moscow State University, Faculty of Computational }\\{Mathematics and Cybernetics, Moscow, Russia }
    \\
    {\it 2. Valiev institute of physics and technology of Russian }\\{Academy of Sciences}
    \\
    }
    \maketitle

    \begin{abstract}
        The singlet state of a system of two two-level atoms changes smoothly, remaining dark, as the Hamiltonian TC is slowly deformed, despite the inapplicability of the adiabatic theorem to this case. In this case, there is a small probability of emission of free photons, which does not depend on the smoothness of the deformation of the Hamiltonian. The effect of spontaneous emission is enhanced by the addition of one more pair of atoms in the singlet state due to the exchange of virtual photons in the cavity. A similar effect was also established for the case when atoms can move between two cavities, but here, on the contrary, with an increase in the number of atoms, the emission decreases. This purely quantum effect must be taken into account in practical manipulations with atomic singlets; however, its weakness testifies, rather, to the stability of dark states and the prospects for their use in information exchange (quantum cryptographic protocols) and as an energy accumulator for nono-devices.
    \end{abstract}

%\newpage
%\tableofcontents
%\newpage

\section{Introduction}

The Jaynes-Cummings model(see \cite{JC}) and its generalization, the Tavis-Cummings model(see \cite{Tav}), describe the dynamics of a group of $n$ two-level atoms in an optical cavity interacting with a single-mode field inside it.

The Hamiltonian of the TCH model has the form
\begin{equation}
\label{TCH}
\begin{array}{ll}
&Н_{TCH}^{RWA}=\sum\limits_i H_{TC,\ i}+\sum\limits_{i<j}\mu_{ij}(a^+_ja_i+a^+_ia_j),\\
 &H_{TC,\ i}=\hbar\w a_i^+a_i+\hbar\w\sum\limits_k\s_{ik}^+\s_{ik}+\sum\limits_k g_k^i(a_i^+\s_{ik}+a_i\s_{ik}^+),
\end{array}
\end{equation}
where the indices $i,j$ refer to the cavity, $k$ to the atom inside it, the atomic relaxation and excitation operators $\s_{ik},s_{ik}^+$ and the field operators $a_i,a_i^+$ have standard view. We will also consider a modification of this model, in which the tunneling of atoms between cavities via special bridges is allowed;in this case, the term $\sum\limits_{i<j}\mu^{at}_{ij}(S_{ik}^+S_{jk}+S_{ik}S_{jk}^+)$ is added to the Hamiltonian \eqref{TCH}, where $S_{jk}$ is the annihilation operator for the $k$ atom in the $j$ cavity, $\mu^{at}_{ij}$ is the tunneling intensity of the $k$ atom between cavities $i,j$. We will call this modification the $TCH^{tun}$ model.

%%%%   %%%%
This model can be extended if we allow the movement of atoms between the cavities. We will consider such a dynamic model below; it is closer to the natural dynamics of atoms in a condensate, in particular, a phenomenon resembling the Frelich effect can be obtained in it (see \cite{Frel}).

%%%%   %%%%
The eigenstates of the Tavis-Cummings Hamiltonian (TC) have a very complicated description (see \cite{Tav}), however, those in which the atomic system cannot interact with the field are of practical value, since it is this interaction that is the main source of decoherence -- the main enemy of quantum computing.

In particular, the so-called dark states of atomic ensembles are important, in which atoms cannot emit energy in the form of photons. Dark states have numerous applications, such as energy storage for nano-devices \cite{2}.
The role of dark states in interatomic interactions was considered in \cite{AL}, in the control of solid-state spins -- in \cite{HS}, in the control of macroscopic quantum systems -- in \cite{LG}; their role in biology -- in light-harvesting complexes was studied in \cite{FH}. Some methods for obtaining dark states can be found in \cite{PB} and in \cite{T}. In a recent paper \cite{1} the dark states of three-level systems were considered. In the paper \cite{O1} the dimension of the dark subspace was established in the Tavis-Cummings model.

The dark subspace in the model with fixed atoms in the TC+RWA model is the kernel of the photon creation operator: $Ker(a^+)$. However, for the $TCH^{tun}$ model with moving atoms, this is no longer the case: an atom can first move and only then emit a photon. Therefore, in the model with moving atoms, the dark subspace (we call it black) will have the form $Ker(a^+U_t)$, where $U_t=exp(-\frac{i}{\hbar}Ht)$ for small $ t>0$.

An important issue is the degree of stability of dark states to a smooth deformation of the Hamiltonian, for example, to the displacement of atoms inside one cavity by optical tweezers, which is expressed in a change in the interaction energy of atoms with the field $g_k^i$, which is the subject of this work.

\section{Deformation of the Hamiltonian for One Cavity}

The interaction energy of an atom with coordinate $x$ inside a cavity of length $L$ and the field of this cavity with wavelength $L/2$ is given by the expression $g=\sqrt{\hbar\w /V}dE(x)$, where $V $ is the effective volume of the cavity, $E(x)=sin(\pi x/L)$. The movement of an atom inside the cavity is reduced to a change in $x$. We will consider the case of such a displacement that $g_1=const$, $g_2\rightarrow 0$. The problem becomes nonstationary and is described by the variable Hamiltonian $H_{TC}(t)$.
The spectrum of the Hamiltonian $TC$ for a diatomic system has no degenerate levels, so the adiabatic theorem applies in this case. 

Numerical simulation shows that the evolution of the solution of the nonstationary Schrödinger equation for such a Hamiltonian, with a sufficiently smooth change in the interaction energy $g_2$, translates the singlet $|s_{12}\rangle=\frac{1}{\sqrt 2}(|01\rangle-|10\rangle)$ 
into a singlet $|s_{12}(t)\rangle=\frac{1}{\sqrt {g_1^2+g_2^2}}(g_1|01\rangle-g_2|10\rangle)$ similar in shape, and if there were no photons in the initial state of the cavity, then they will not appear during the entire evolution, which indicates the stability of one singlet to smooth deformation of the Hamiltonian.

\begin{figure}[H]
    \centering
    \includegraphics[scale=0.6]{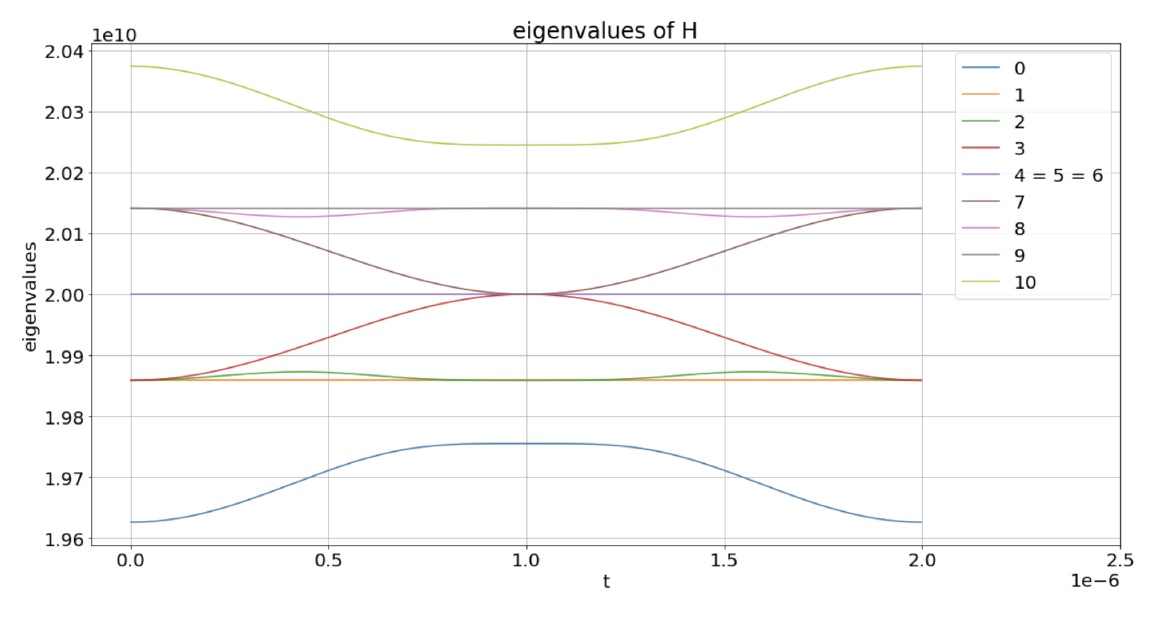} 
    \caption{Spectral lines dependent from time of the deformable Hamiltonian TC. There is numeration from null to tenth eigenstates with increasing energy.}
    \label{fig:2.1}
\end{figure}

However, for two singlets in the same cavity, the situation will be completely different. For a 4 atomic quartet, the spectrum of the deformable Hamiltonian TC has the form shown in the Figure \ref{fig:2.1}; the degeneracy of the levels corresponding to dark states makes it impossible to apply the adiabatic theorem.  The fourth level on the picture will be initial state dark state $|s_{12,34}\rangle$, but degenerate fifth and sixth levels contain nonzero amplitude of basis state with explicit 2 photons ($|2\rangle_{ph}|0000\rangle_{at}$).

\section{Stability of Dark States for a Model with Displacement of Atoms}

Let us consider the possibility of tunneling atoms between cavities -- the $TCH^{tun}$ model. We assume that the graph ${\cal G}$ of cavities with edges in the form of tunnel bridges for atoms between them is connected.
For a diatomic system, dark states of its Hamiltonian will exist if and only if the graph ${\cal G}$ is even, that is, any cycle in it contains an even number of vertices. In this case, the dark state of 2 atoms is the only one and has the form $|s_{\cal G}\rangle=\sum\limits_i(-1)^{d(i)}|s^i\rangle$, where $d(i)$ is the length of the path connecting cavity $i$ to cavity $0$, and $|s^i\rangle= |s^i_{12}\rangle$ is a singlet in cavity $i$.

%%%%   %%%%
The phenomenon of dark states for moving atoms is similar to the Frelich effect \cite{Frel}, when the initial excitation of high vibrational modes passes into a stable excited state of atomic vibrations. In our model, the dark state $|s_{\cal G}\rangle$ serves as a skeleton of such a stable state: the dark dynamic state does not interact with the field, and therefore any excitation with subsequent relaxation only increases this skeleton.

%%%%   %%%%

Let the graph ${\cal G}$ consist of only 2 cavities, so that $|s_{\cal G}\rangle=|s^1\rangle-|s^2\rangle$. We will assume that this process is carried out in such a way that the energy of interaction with the field for the displaced atom is conserved; this can be ensured by arranging tunneling channels separately for each atom. Let $g_1^1=g_1^2=const,\ g_1^2=g_2^2\rightarrow 0$ be a slow deformation of the Hamiltonian.

%%%%%%%%%%%%%%%%%%%%%%%%%%%%%%

\section{Probability of Free Photons Appearing in Optical Cavity}

To make $g_2(t)$(and $g_4(t)$) change slowly enough, use the following function:
\begin{figure}[H]
\centering
\includegraphics[scale=0.5]{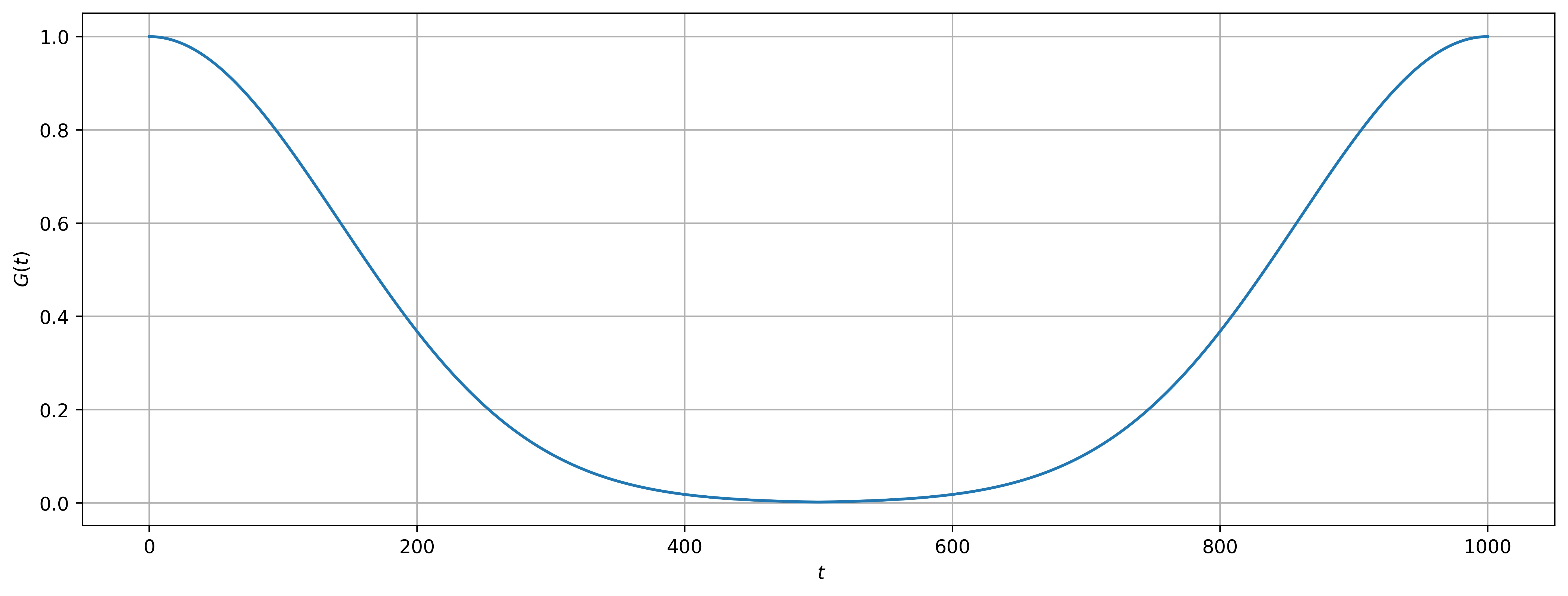} 
\caption{$G(t)$}
\label{fig:2}
\end{figure}

\begin{eqnarray}
    G(t)=
    \begin{cases}
        e^{-(5 \frac{t}{T})^2} & 0\leq t\leq 0.5T\\
        e^{-(5 \frac{T-t}{T})^2} &  0.5T\leq t\leq T 
    \end{cases}
    \label{formula2}
\end{eqnarray}
where $t\in[0,T],\ \Delta t=0.001,\ T=10^3,\ f'(0.5T-0)=f'(0.5T+0)\approx 0,\ g_2(t)=g_1 G(t)$.

Schrödinger equation:
\begin{eqnarray}
    \begin{cases}
        i\hbar|\dot{\psi}(t)\rangle=H(t)|\psi(t)\rangle & 0\leq t\leq T\\
       |\psi(0)=|s_{12}\rangle 
    \end{cases}
\end{eqnarray}

Iterative calculation:
$|\psi(t+\Delta t)\rangle=e^{-iH(t)\Delta t}|\psi(t)\rangle$
~\\

1.Suppose there are two atoms in the cavity. We set the initial state as $|s_{12}\rangle=\frac{1}{\sqrt 2}(|01\rangle-|10\rangle)$.
(If there are 4 atoms in a cavity, then the initial state is:$|\psi(0)\rangle=|s_{12}\rangle|s_{34}\rangle$)

\begin{figure}[H]
\centering
\includegraphics[scale=0.5]{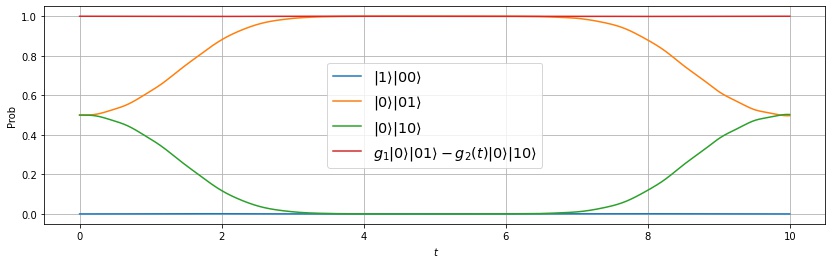} 
\caption{2 atoms,1 cavity,the probability of intracavity states.$t\in[0,T']$ $T'=0.1T$}
\label{fig:3}
\end{figure}

\begin{figure}[H]
\centering
\includegraphics[scale=0.5]{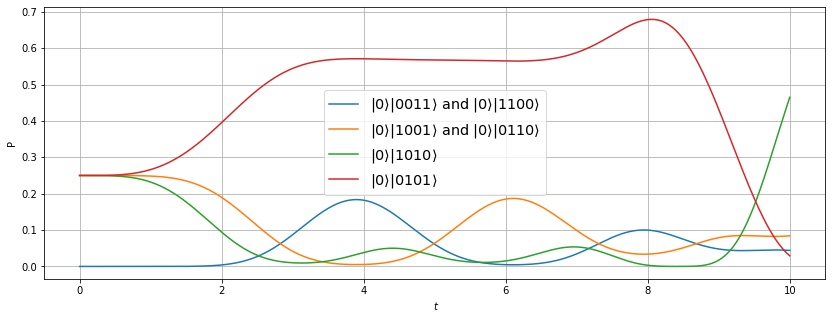} 
\caption{4 atoms, 1 cavity, initial state:$|s_{12,34}\rangle=|s_{12}\rangle|s_{34}\rangle$ $t\in[0,T']$ $T'=0.1T$}
\label{fig:4}
\end{figure}

In the case of two atoms, the state in the cavity returns to the initial state as $g_2(t)$ returns to its initial value.
However, as the number of atoms increases, when the function $G(t)$ changes rapidly(In Figure \ref{fig:4}, $G(t)$ changes 10 times faster than the original), the state in the resonator does not return to the initial state (Figure \ref{fig:4}).

Therefore, we carry out simulation experiments using a slowly varying function $G(t)$(Figure \ref{fig:2}).\\

2. The probability of the appearance of at least one free photon.
In the presence of only two atoms in the resonator, the probability of a free photon appearing is as follows:
\begin{figure}[H]
\centering
\includegraphics[scale=0.5]{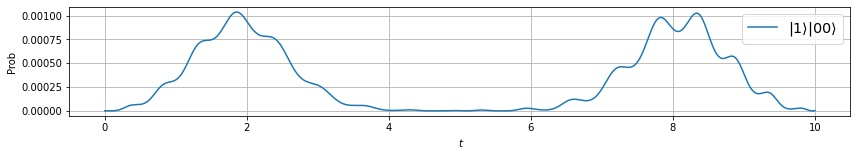} 
\caption{2 atoms, 1 cavity, the probability of the appearance of free photons ($|1\rangle|00\rangle$)}
\label{fig:5}
\end{figure}

In the case of 4 atoms in the resonator, we assume that the initial state is $|s_{12,34}\rangle=|s_{12}\rangle|s_{34}\rangle$.

\begin{figure}[H]
\centering
\includegraphics[scale=0.5]{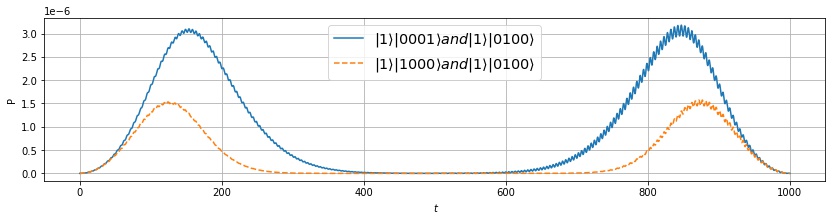} 
\caption{4 atoms, 1 cavity, probability of $|1\rangle|0001\rangle,\ |1\rangle|0010\rangle,\ |1\rangle|0100\rangle$, and $|1\rangle|1000\rangle$ ($g_2(t)=g_4(t)=G(t)$).}
\label{fig:6}
\end{figure}

These two graphs are probability curves for the appearance of only one photon in the resonator. We see that whether there are only 2 atoms or 4 atoms in the resonator, their probability curves show two humps.

\begin{figure}[H]
\centering
\includegraphics[scale=0.5]{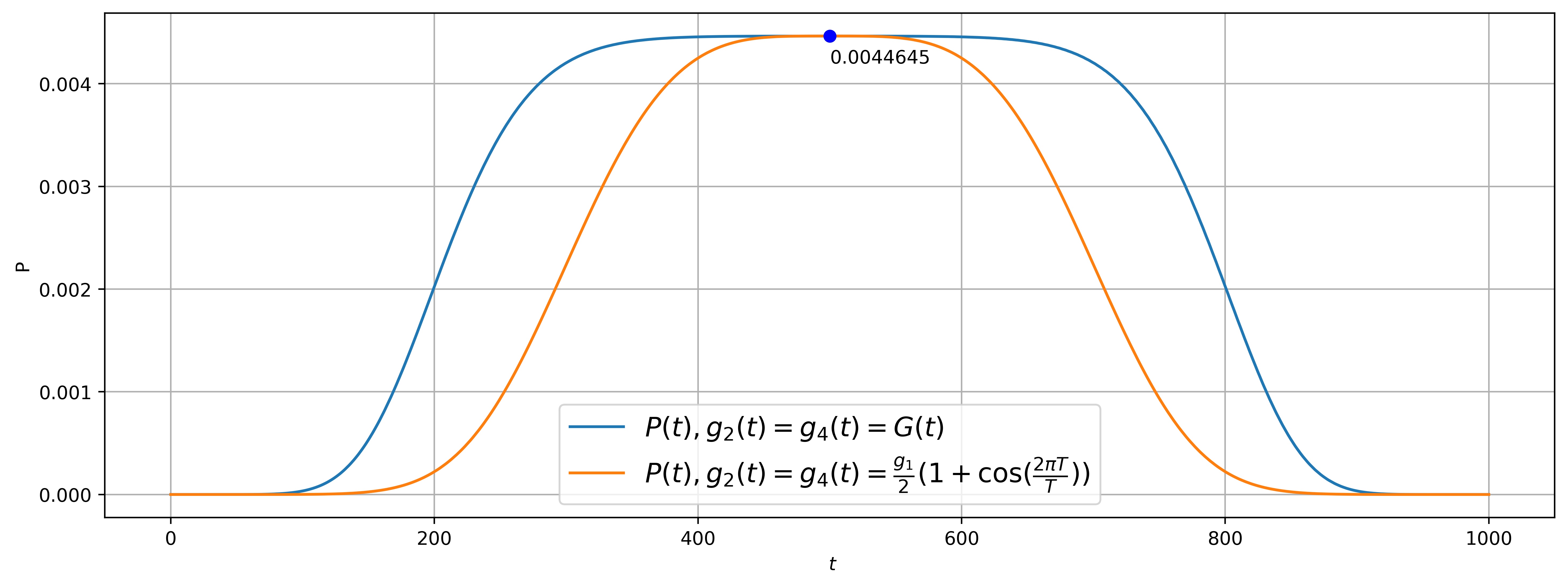} 
\caption{4 atoms, 1 cavity, probability that there will be at least one explicit photon ($g_2(t)=g_4(t)=G(t)$).}
\label{fig:7}
\end{figure}

$P(t)$ is the probability that there will be at least one explicit photon in the cavity at time t.

$P(t)=\langle 2_{ph}|\rho_{ph}(t)|2_{ph}\rangle + \langle1_{ph}|\rho_{ph}(t)|1_{ph}\rangle$

$\rho_{ph}=tr_{ph}(\rho(t))$-relative field density matrix;

$\rho(t)=|\Psi\rangle\langle\Psi|$ is the density matrix of the entire state of the field and atoms during slow evolution. 

Figure \ref{fig:7} is the case where 4 atoms are in one cavity. Because the probability of occurrence of 2 photons (the maximum value is about 0.004) is much greater than the probability of only one photon (the maximum value is about $3\times 10^{-6}$), the image basically presents the probability of occurrence of two photons, and the curve presents single hump.

Comparison of Figure \ref{fig:5} with Figure \ref{fig:7} shows that the presence of an additional pair of atoms in the singlet state increases the effect of spontaneous photon emission. 

We use other functions $g_2(t)=\frac{g_1}{2}(1+\cos(\frac{2\pi t}{T}))$ and, by choosing different smooth deformations, we find the result of the maximum amplitude of free photons (at $t=0.5T$, $g_2(t)=g_4(t)=0$) hardly changes.

The two images have nearly identical maxima. We have reason to think that it is not a coincidence that free photons exist.\\

3.Method for assessing whether the state in an optical cavity remains dark during changes in $g_i(t)$.

Because the probability of photons appearing in the optical cavity is small, we need to verify whether this is an iterative error or is caused by the destruction of the dark state during the changes of $g_2(t)$ and $g_4(t)$. We use the following method to check whether the state inside the optical cavity remains dark all the time.

\begin{equation*}
	\overline{\sigma}(t)=g_1 \sigma _1+g_2 (t) \sigma _2+g_3 \sigma _3+g_4 (t) \sigma _4
\end{equation*}

After simplifying the matrix $\overline{\sigma}(t)$, we obtain $rank(\overline{\sigma}(t))=6$. Therefore, the solution space $\overline{\sigma}(t)|s\rangle=0$ has 6 bases:
\begin{eqnarray}
    \begin{cases}
       |s_1(t)\rangle=g_1 g_3|0101\rangle+g_2 (t)g_4 (t)|1010\rangle-g_1 g_4 (t)|0110\rangle-g_2 (t) g_3|1001\rangle\\

|s_2(t)\rangle=-g_1 g_3|0101\rangle-g_2 (t)g_4 (t)|1010\rangle-g1g_2 (t)|0011\rangle+g_3g_4 (t)|1001\rangle\\

|s_3(t)\rangle=-g_1|0100\rangle-g_2 (t)|1000\rangle\\

|s_4(t)\rangle=-g_1|0001\rangle-g_4 (t)|1000\rangle\\

|s_5(t)\rangle=-g_1|0010\rangle-g_3 (t)|1000\rangle\\

|s_6(t)\rangle=|0000\rangle\\
    \end{cases}
\end{eqnarray}

Since the initial state is $|s_{12,34}\rangle$, then if $|\psi (t)\rangle$ remains in the dark state and no free photons appear in the resonator, then the state in the cavity can be represented as: $ |\psi(t)\rangle=\alpha|s_1(t)\rangle+\beta|s_2(t)\rangle$. This means that the ratio of amplitude $|0101\rangle$ and amplitude $|1010\rangle$ is: $\frac{g_1 g_3}{g_2 (t)g_4 (t)}$.We set the function $\lambda(x)$ -- the amplitude of $x$.
So we should have $g_2 (t)g_4 (t)\lambda(|0101\rangle)-g_1 g_3\lambda(|1010\rangle)=0$ if the state in the cavity remains dark all the time.

\begin{figure}[H]
\centering
\includegraphics[scale=0.5]{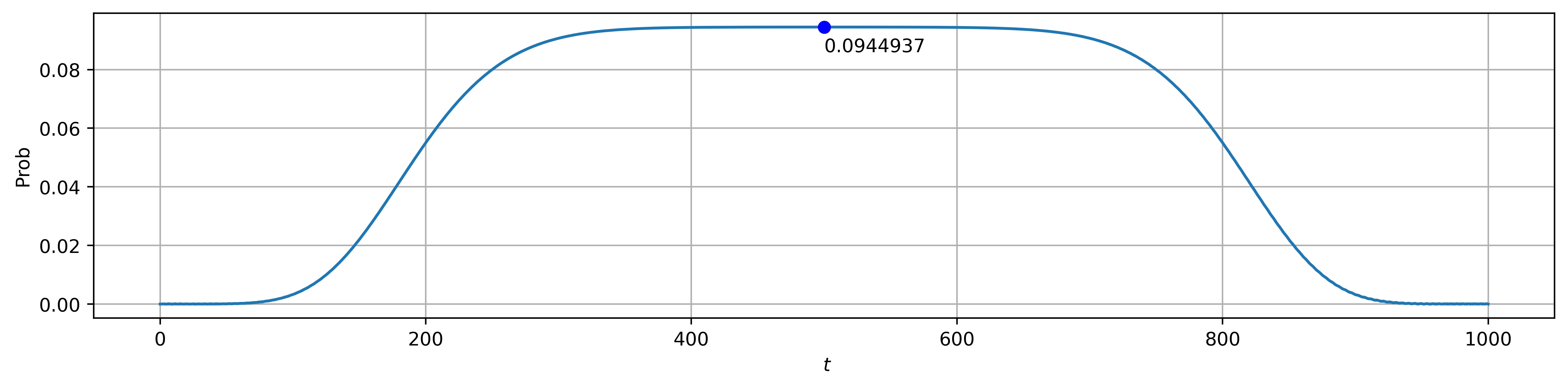} 
\caption{4 atoms,1 cavity,$g_2 (t)g_4 (t)\lambda(|0101\rangle)-g_1 g_3\lambda(|1010\rangle)$}
\label{fig:8}
\end{figure}

From here we see that $g_2 (t)g_4 (t)\lambda(|0101\rangle)-g_1 g_3\lambda(|1010\rangle)$ does not always remain 0. Even at $t=0.5T$ it is much larger than 0. This means that $|\psi(t)\rangle$ cannot be represented by a dark state. So we are sure that there are free photons in the optical cavity.\\

%%%%%%%%%%%%%%%%%%%%%%%%%%%
    \section{Probability of Free Photons in the Resonator of the Modified Model $TCH^{tun}$}
    
    In chapter 1, we presented the $TCH^{tun}$ model, in which the tunneling of atoms between cavities via special bridges is allowed.
    
    Thus, the basic state will look like:

	\begin{equation*}	
		|n_1,n_2,...,n_k\rangle_{ph}\,|at_1 state,\,at_1position\rangle|at_2state,\, at_2positon\rangle...|at_nstate,\,at_nposition\rangle.
	\end{equation*}

    where
    
    \begin{itemize}
		\item $|n_1, n_2, ..., n_k\rangle_{ph}$ -- as usual, the number of photons in the cavities $1, 2,..., k$.
		\item $| at_istate,at_iposition\rangle$ characterizes the atom $i$.
		\item $at_istate\in \{0,1 ,..., d\}$.
		\item $at_iposition\in \{ 1,2, ..., k\}$ -- the cavity in which it is located.
	\end{itemize}
	
	Next, we solve the Schrödinger equation to study the change in the initial state and the probability of emitting a photon as the strength of the atom and field change.
	
	\subsection{Two atoms, two cavities}
    Two cavities, two atoms in a singlet state. One stands still, while the other has a transition bridge that we slowly move towards the mirror (that is, it has $g^1_2 = g^2_2 \to 0$), that is, in both the first and second cavity, the force of interaction with the field slowly tends to zero. Let the atoms initially have the same $g$ in both cavities. That is, $g_{1}^{1}=g_{1}^{2}=g_{2}^{1}=g_{2}^{2} = g = const$.
     
    We use the same function $G(t)$ (Figure \ref{fig:2}) as in the previous chapter. That is, the force acting on the atom $j$ in the cavity $i$ has the function $g_{j}^{i}(t) = g \times G(t)$ .
     
    Initial state -- black state $|s_1\rangle - |s_2\rangle$
    \begin{figure}[H]
        \centering
        \includegraphics[width=6in]{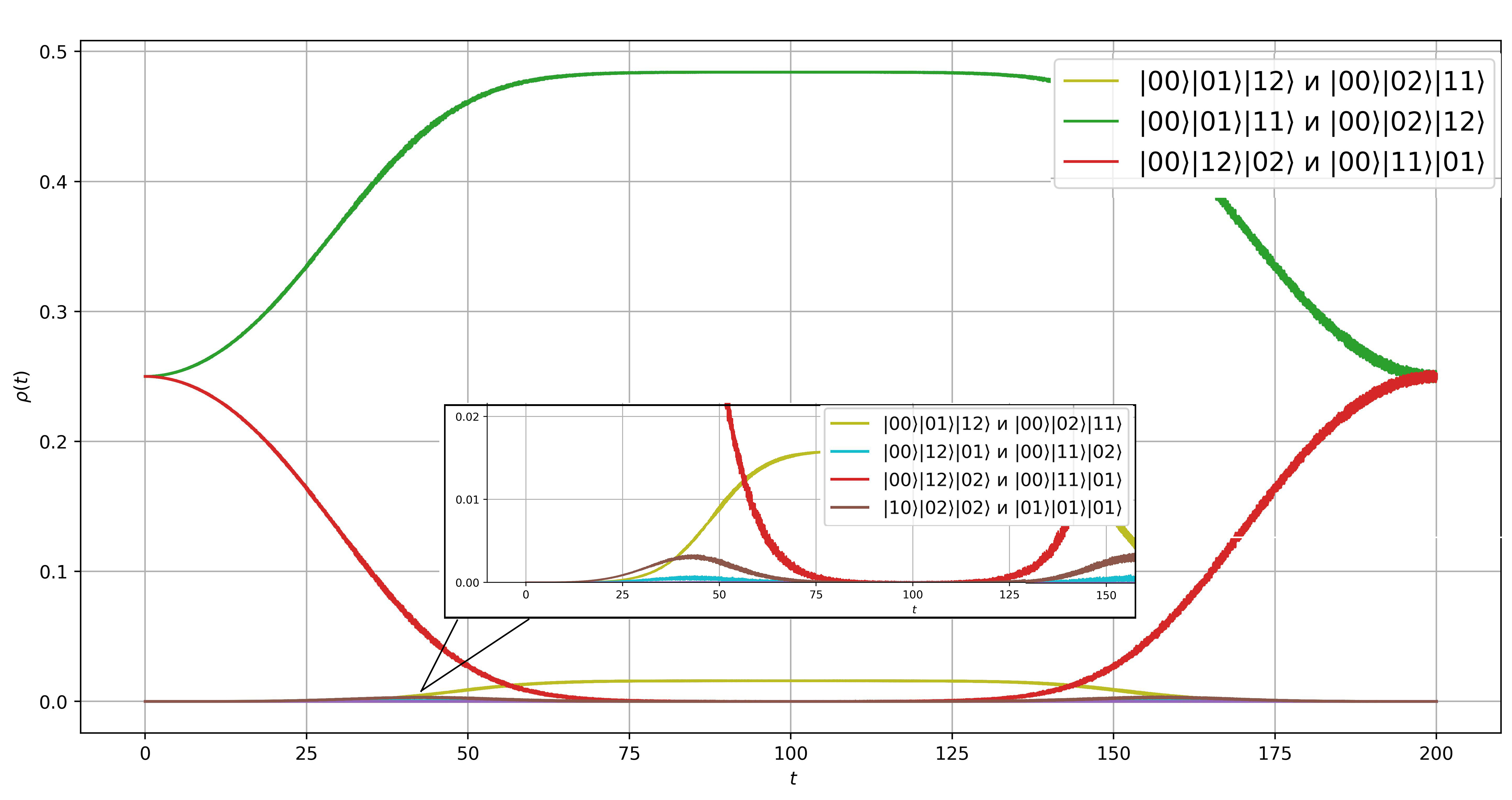}
        \caption{Two atoms, two cavities, photons and atoms can move between cavities. $g^1_2 = g^2_2 \to 0$ slowly adiabatically. Two atoms in a singlet state: $\Psi(0)=|s_1\rangle - |s_2\rangle$.}
        \label{fig:imagerho_a_20000.jpg}
    \end{figure}
    
    In Figure \ref{fig:imagerho_a_20000.jpg} we see that when the second atom $g_2$ decreases, two atoms can be in different cavities.

    \begin{figure}[H]
        \centering
        \includegraphics[width=6in]{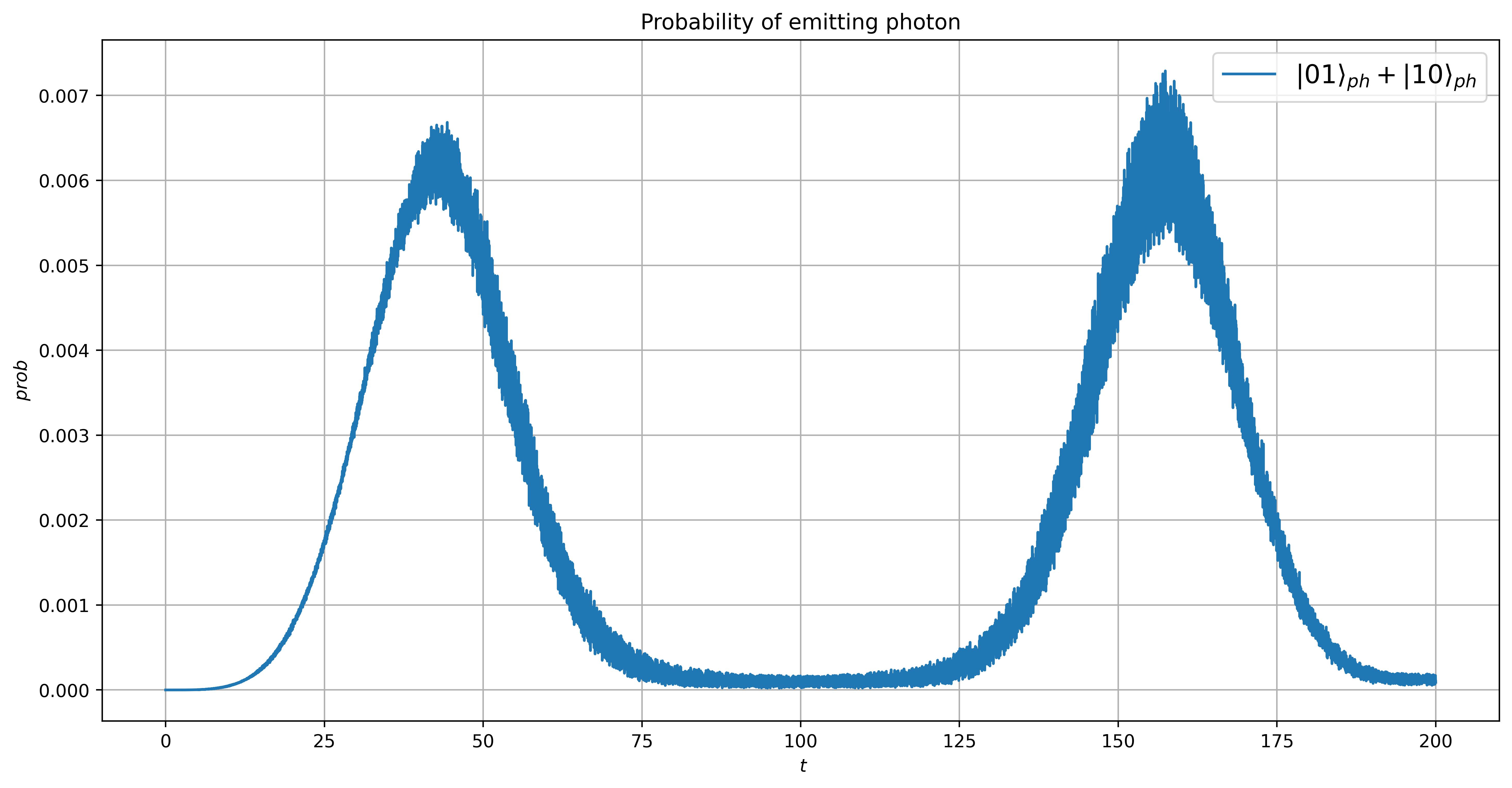}
        \caption{Two atoms, two cavities, photons and atoms can move between cavities. $g^1_2 = g^2_2 \to 0$ slowly adiabatically. Probability of photon emission.}
        \label{fig:image_emitting_photon_a_20000.jpg}
    \end{figure}
    
    Probability function of emission of at least one photon (Figure \ref{fig:image_emitting_photon_a_20000.jpg}) values: $prob(t) = \langle01_{ph}|\rho(t)|01_{ph}\rangle + \langle10_{ph}|\rho(t)|10_{ph}\rangle$, where $\rho(t)=|\Psi\rangle\langle\Psi|$.
    
    \subsection{The amplitude of an atom transition between cavities changes}

    And if $g_{1,2}^i=const$, but the amplitude of the transition of the atom between the cavities changes. We use the same function $G(t)$(Figure \ref{fig:2}). That is, the change in the intensity of the movement of atoms between the cavities has the function $R(t) = r\times G(t)$.
    
    The result obtained is close to a straight line. That is, the initial state $|s_1\rangle-|s_2\rangle$ was approximately stable after the change in time $t$.
    
\begin{figure}[H]
        \centering
        \includegraphics[width=6in]{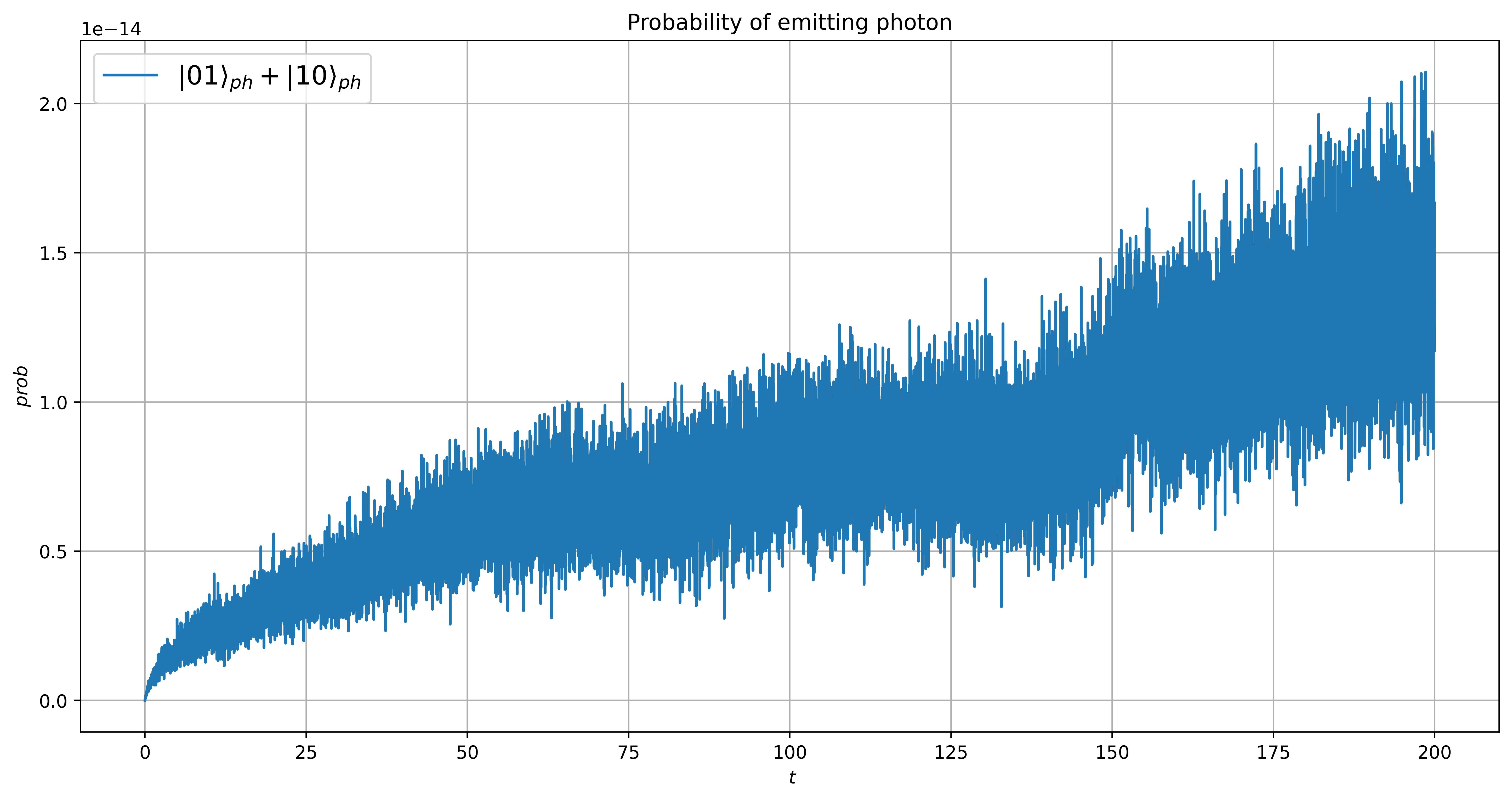}
        \caption{Changing the intensity of movement of atoms between cavities. Probability of photon emission.}
        \label{fig:image_emitting_photon_b_20000.jpg}
    \end{figure}

    But the probability of emitting a photon still exists, although the probability is very small, about $10^{-14}$, as shown in Figure \ref{fig:image_emitting_photon_b_20000.jpg}.

    \subsection{Four atoms, two cavities}
    %\subsection{Adiabatic system of four two-level atoms in two cavities}
    
    Two cavities, four atoms in a singlet state. $g_2^1=g_2^2=g_4^1=g_4^2 \to 0$, that is, both in the first and second cavity, the force of interaction of the second and fourth atoms with the field slowly tends to zero. Let the atoms initially have the same $g$ in both cavities.
    
    We use the same function $G(t)$ (Figure \ref{fig:2}) as the change $g$ of the second and fourth atoms in the cavity.
    
    Initial state:
    
    \begin{equation*}
        |\Psi(0)\rangle = (|s_1^{12}\rangle - |s_2^{12}\rangle)\otimes (|s_1^{34}\rangle - |s_2^{34}\rangle)
    \end{equation*}

    \begin{figure}[H]
    \centering
        \includegraphics[width=6in]{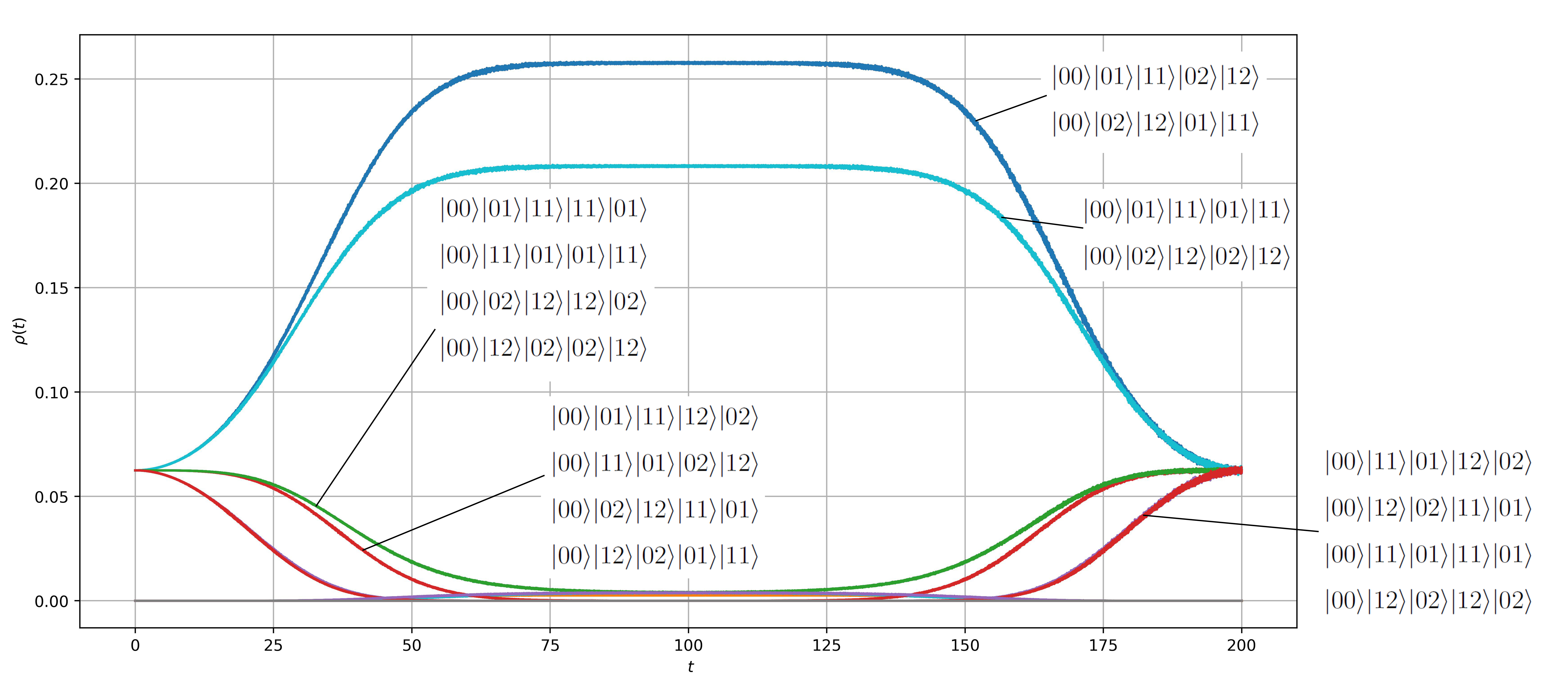}
        \caption{Four atoms, two cavities, photons and atoms can move between cavities. $g_2^1,g_2^2,g_4^1,g_4^2\to 0$ slowly adiabatically. Initial state : $\Psi(0)=(|s_1^{12}\rangle - |s_2^{12}\rangle)\otimes (|s_1^{34}\rangle - |s_2^{34}\rangle) $.}
        \label{fig:imagerho_c_20000.jpg}
    \end{figure}

\begin{figure}[H]
        \centering
        \includegraphics[width=6in]{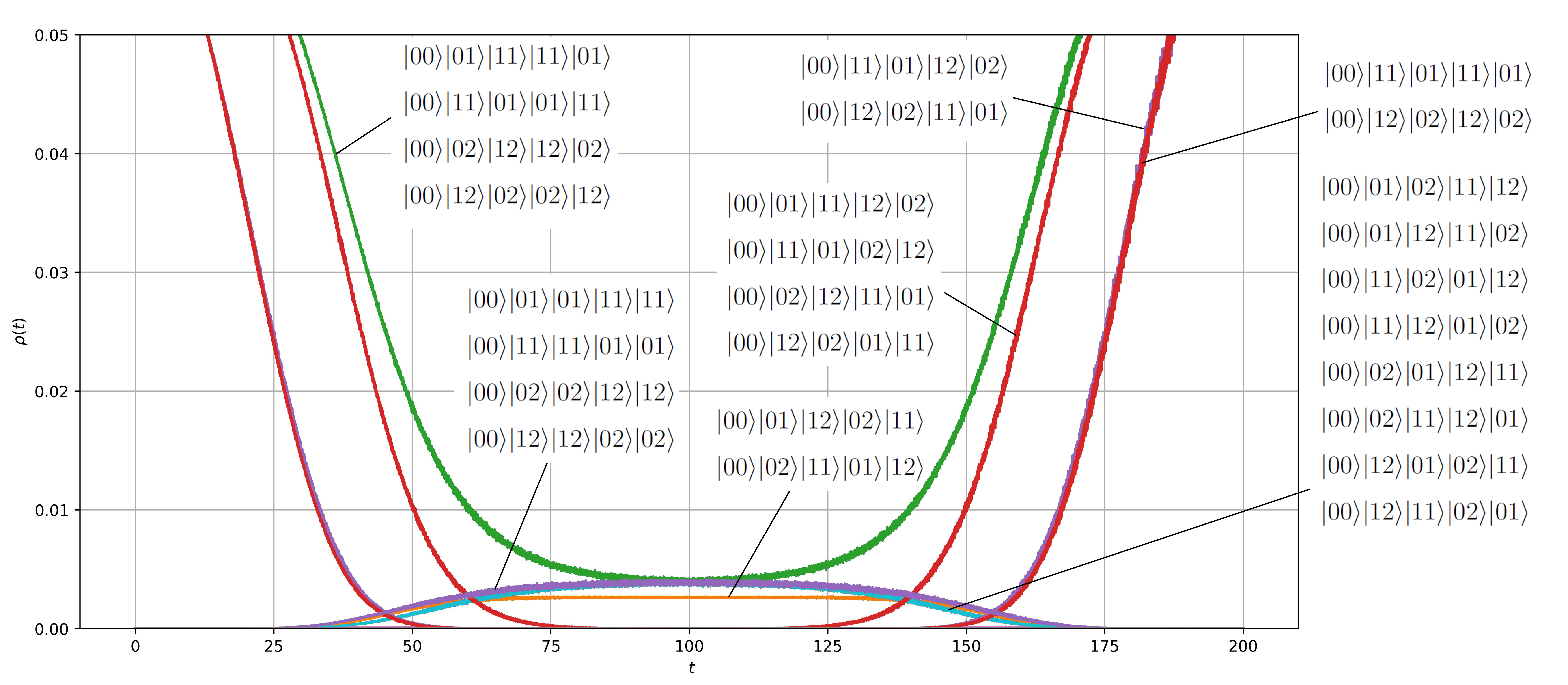}
        \caption{Four atoms, two cavities, photons and atoms can move between cavities. $g_2^1,g_2^2,g_4^1,g_4^2\to 0$ slowly adiabatically. Initial state : $\Psi(0)=(|s_1^{12}\rangle - |s_2^{12}\rangle)\otimes (|s_1^{34}\rangle - |s_2^{34}\rangle) $.}
        \label{fig:imagerho_c_20000_limy_0.05.jpg}
    \end{figure}

    We enlarge the figure and can see that as the second and fourth atoms $g_2^i,g_4^i$ decrease, various combinations of four atoms appear in the cavity.(Figure \ref{fig:imagerho_c_20000_limy_0.05.jpg})
    
    \begin{figure}[H]
        \centering
        \includegraphics[width=6in]{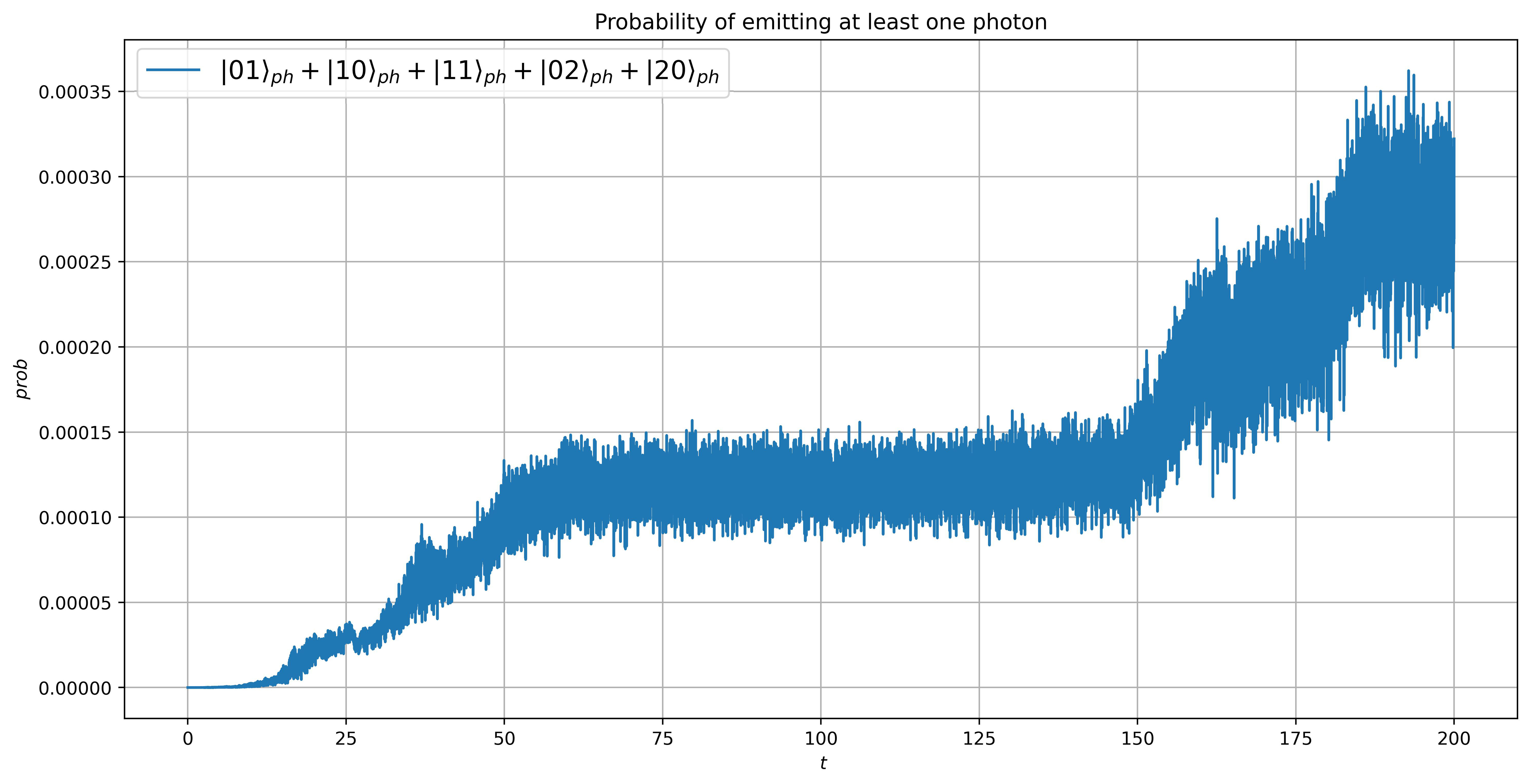}
        \caption{Probability of at least one photon being emitted.}
        \label{fig:image_emitting_photon_c_20000.jpg}
    \end{figure}
    
    Probability function of emission of at least one photon (Figure \ref{fig:image_emitting_photon_c_20000.jpg}) values: $prob(t) = \langle01_{ph}|\rho(t)|01_{ph}\rangle + \langle10_{ph}|\rho(t)|10_{ph}\rangle + \langle11_{ph}|\rho(t)|11_{ph}\rangle + \langle02_{ph}|\rho(t)| 02_{ph}\rangle + \langle20_{ph}|\rho(t)|20_{ph}\rangle$, where $\rho(t)=|\Psi\rangle\langle\Psi|$.
    
    \subsection{Parameters used for calculation}
    \begin{equation}
        \begin{aligned}
            &\hbar = 1.054571817 * 10^{-27} & \Delta t = 0.01 \\
            &\mu_{12} = 9 * 10^{7} &\mu_{12}^{at} = 8 * 10^{7}\\
            & g = 2 * 10^{8} & \omega = 10^{7}
        \end{aligned}\label{fig:d2parameter}
    \end{equation}

%%%%%%%%%%%%%%%%%%%%%%%%%%%

\section{Conclusions}

1. A paradoxical effect of spontaneous emission of photons by the dark state of a diatomic system in an optical cavity with an arbitrarily smooth deformation of the Tavis-Cummings Hamiltonian has been established, which takes place despite the inapplicability of the adiabatic theorem to this case. The deformation of the Hamiltonian physically means the slow movement of atoms in the cavity with the help of optical tweezers. This effect bears some resemblance to the shift of the electron wave front in the vicinity of the solenoid (the Aharonov-Bohm effect).

2. It is shown that this effect is greatly enhanced for 4 atoms in the dark state. This is a purely quantum effect of the influence of singlets on each other, while separately they do not interact with the field.

3. A similar effect also takes place under the condition that atoms move between cavities; however, in this case, the addition of atoms does not enhance the effect, but, on the contrary, weakens it. The effect of spontaneous emission upon deformation of the Hamiltonian is stronger in the case of a single field in one cavity than for a separated field in two cavities. The mechanism of the effect is associated with the exchange of virtual photons between the singlet states of pairs of atoms, which become real upon deformation of the Hamiltonian. When changing the amplitude of the transition of atoms between the cavities while maintaining the strength of interaction with the field for each atom, the effect of spontaneous emission is negligible.

The effect of spontaneous photon emission by dark states must be taken into account when manipulating atoms in practical applications of dark states, however, the very low probability of photon emission during smooth deformation of the Hamiltonian indicates the high stability of such states and the prospects for using these states to store information or as energy accumulators for nanosystems.

\section{Acknowledgements}

The work was carried out at the Moscow Center for Fundamental and Applied Mathematics.

\end{document}